\documentclass[preprint,showpacs,superscriptaddress,preprintnumbers,amsmath,amssymb]{revtex4}

\usepackage{graphicx, color}
\usepackage{dcolumn}
\usepackage{bm}
\usepackage{amsmath}

\newcommand\p{\partial}

\usepackage[normalem]{ulem}  

\renewcommand\sout{\bgroup \color{red} \ULdepth=-.5ex \ULset}

\begin{document}

\title{
Quantized chiral magnetic current\\
from reconnections of magnetic flux
}

\author{Yuji~Hirono}
\affiliation{Department of Physics, Brookhaven National Laboratory,
Upton, New York 11973-5000}
\email{yhirono@bnl.gov}
\author{Dmitri~E.~Kharzeev}
\affiliation{Department of Physics, Brookhaven National Laboratory, Upton, New York 11973-5000}
\affiliation{Department of Physics and Astronomy, Stony Brook University, Stony Brook,
 New York 11794-3800, USA}
\affiliation{RIKEN-BNL Research Center, Brookhaven National Laboratory, Upton, New York 11973-5000}
\author{Yi Yin}
\affiliation{Department of Physics,
Brookhaven National Laboratory, Upton, New York 11973-5000}

\date{\today}

\begin{abstract}
We introduce a new mechanism for the chiral magnetic effect that does
not require an initial chirality imbalance.  The chiral magnetic
current is generated by reconnections of magnetic flux that change
the magnetic helicity of the system. The resulting current is entirely
determined by the change of magnetic helicity, and it is quantized.
\end{abstract}

\maketitle

The chiral magnetic effect (CME) is the generation of electric current
induced by the chirality imbalance in the presence of a magnetic field
\cite{Kharzeev:2004ey}; see Ref.~\cite{Kharzeev:2013ffa} for a review and
references to related works. It is a macroscopic manifestation of the
chiral anomaly \cite{Adler:1969gk,Bell:1969ts}. In most of the previous
works reviewed in
Refs.~\cite{Kharzeev:2013ffa,Kharzeev:2015kna,Kharzeev:2015znc}, the
chirality
imbalance is assumed to be generated by a topologically nontrivial
background -- for example, by parallel external electric and magnetic
fields, or by the non-Abelian sphaleron transitions. In fact, recent
experimental observations of the CME current in Dirac semimetals 
\cite{Li:2014bha,Xiong2015,Huang2015} utilized parallel electric and
magnetic fields.
The magnetic field is usually introduced as an external background as
well, even though a number of studies have addressed the role of the CME and
anomaly-induced phenomena in the generation of magnetic helicity 
\cite{Joyce:1997uy,
Boyarsky:2011uy,Tashiro:2012mf,Tuchin:2014iua,Manuel:2015zpa}.
For example, recently it was pointed out that the CME leads to a
self-similar inverse cascade of magnetic helicity
\cite{Hirono:2015rla,Yamamoto:2016xtu}.

\vskip0.3cm
In this Letter, we do not assume the presence of a chirality imbalance
generated by an external topological background -- instead, we consider
the chirality associated with the topology of the magnetic flux
itself. Indeed, in the absence of magnetic monopoles, the lines of
magnetic field have to be closed. For example, the field lines of a
solenoid form an ``unknot.'' However, the topology of the magnetic flux can be
more complicated -- specifically, the magnetic flux can form a chiral
knot. Magnetic reconnections can change the chirality of this knot and
induce the imbalance of chirality in the system. Can this imbalance of
chirality lead to the generation of the chiral magnetic current? We will
show that the answer to this question is positive. The corresponding
chiral magnetic current is quantized, and is completely determined
by the knot invariants. 

\vskip0.3cm

\begin{figure}[htbp]
\centering
\includegraphics[width=0.7\textwidth]{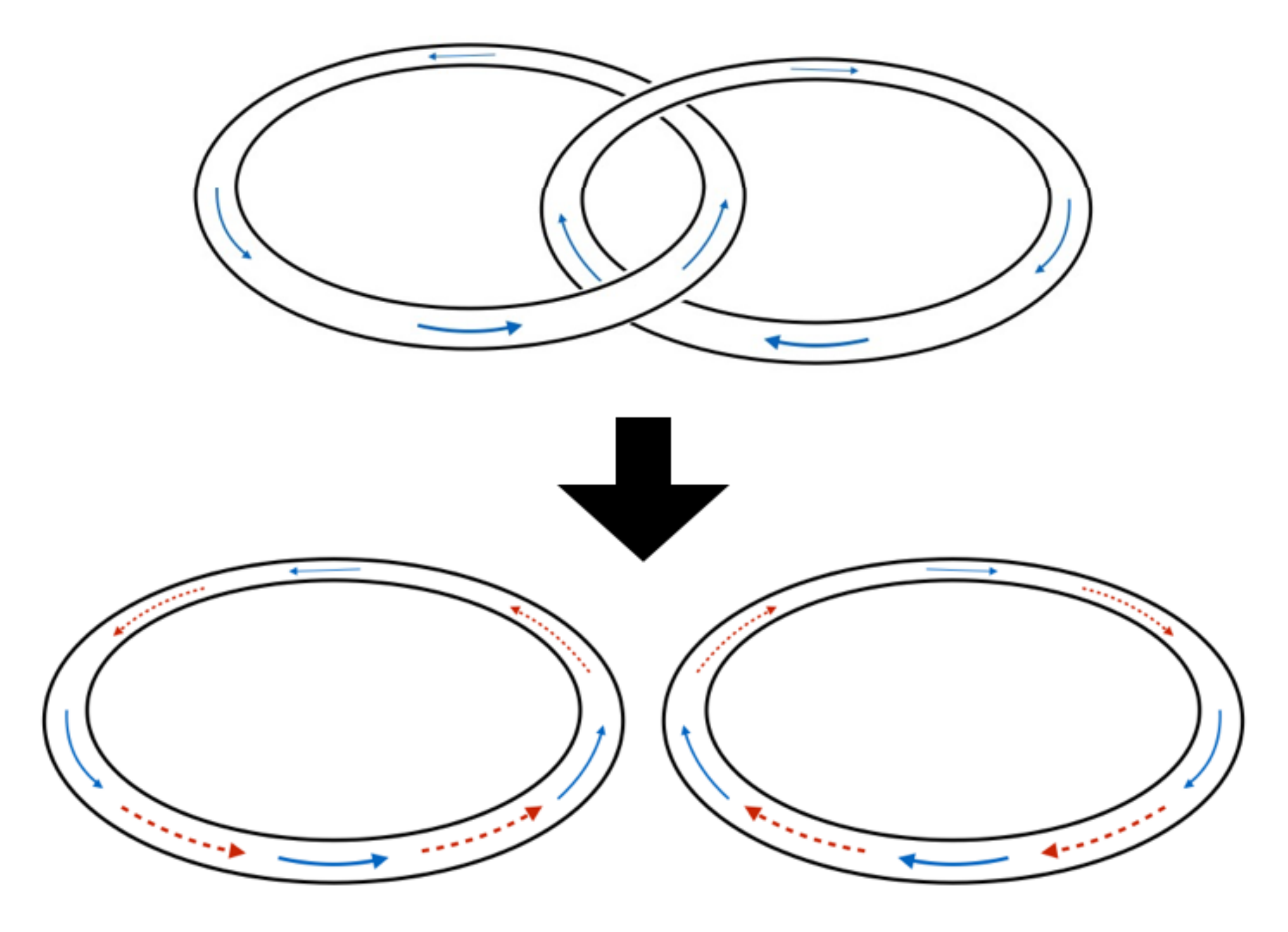}
\caption{
\label{fig:unlink1}
 (Color online)
 Current generation associated with the unlinking of a simple link of two
 flux tubes.
 The solid arrows denote the directions of the magnetic field,
 and the dotted arrows indicate the directions of the generated CME currents. 
 }
\end{figure}

Our main result is the following formula for the generated current $\Delta \bm J$ 
along the loops $C_i$ of the magnetic flux  in terms of the change 
$\Delta \mathcal H$ of the magnetic helicity,
which is a topological measure of the knot (to be defined below):
\begin{equation}
\sum_{i} \oint_{C_i} \Delta \bm J \cdot d \bm x = - \frac{e^3}{2 \pi^2}
\Delta \mathcal H ,
\label{main_res}
\end{equation}
where $e$ is the electric charge. Since $\Delta \mathcal H$ is
an integer number times the flux squared,
the CME current resulting from  reconnections of the magnetic flux is
quantized.
The process illustrated in Fig.~\ref{fig:unlink1} shows the simplest
realization of such currents. 
This unlinking of a link involves the topology change of the magnetic fluxes, 
which leads to the generation of CME currents (indicated by dotted arrows) on both tubes.
The amount of integrated current over the tubes is given by the helicity
change during the process, as quantified by Eq.~(\ref{main_res}). 

\vskip0.3cm

%

Let us now present the derivation of Eq.~(\ref{main_res}). Consider a set of
closed tubes of the magnetic flux, in the presence of massless fermions. 
Repeated reconnections performed on this set will yield a
topologically nontrivial structure containing links and knots of
the magnetic flux; see the upper figure of Fig. \ref{fig:unlink1} for a
simple example.
A nontrivial topology can also be introduced by the twisting of a flux
tube 
\cite{berger1984topological,moffatt1990energy,Ricca1992,moffatt1992helicity}. 
A link $\mathcal K$ of $N$ knots of the magnetic flux tubes can be characterized by 
{\it magnetic helicity}  $\mathcal H$ (an Abelian Chern-Simons 3-form) that
can be decomposed as
\begin{equation}
 \mathcal H (\mathcal K)
  = \sum_{i} \mathcal S_i \varphi^2_i
   + 2 \sum_{i,j} \mathcal L_{ij} \varphi_i \varphi_j , 
\end{equation}
where
$\varphi_i$ is the magnetic flux of the $i$th closed tube, 
$\mathcal S_i$ is the C\u{a}lug\u{a}reanu-White self-linking number,
and $\mathcal L_{ij}$ is the Gauss linking number 
\cite{moffatt1990energy,Ricca1992,moffatt1992helicity}.

\begin{figure}[htbp]
\centering
\includegraphics[width=0.5\textwidth]{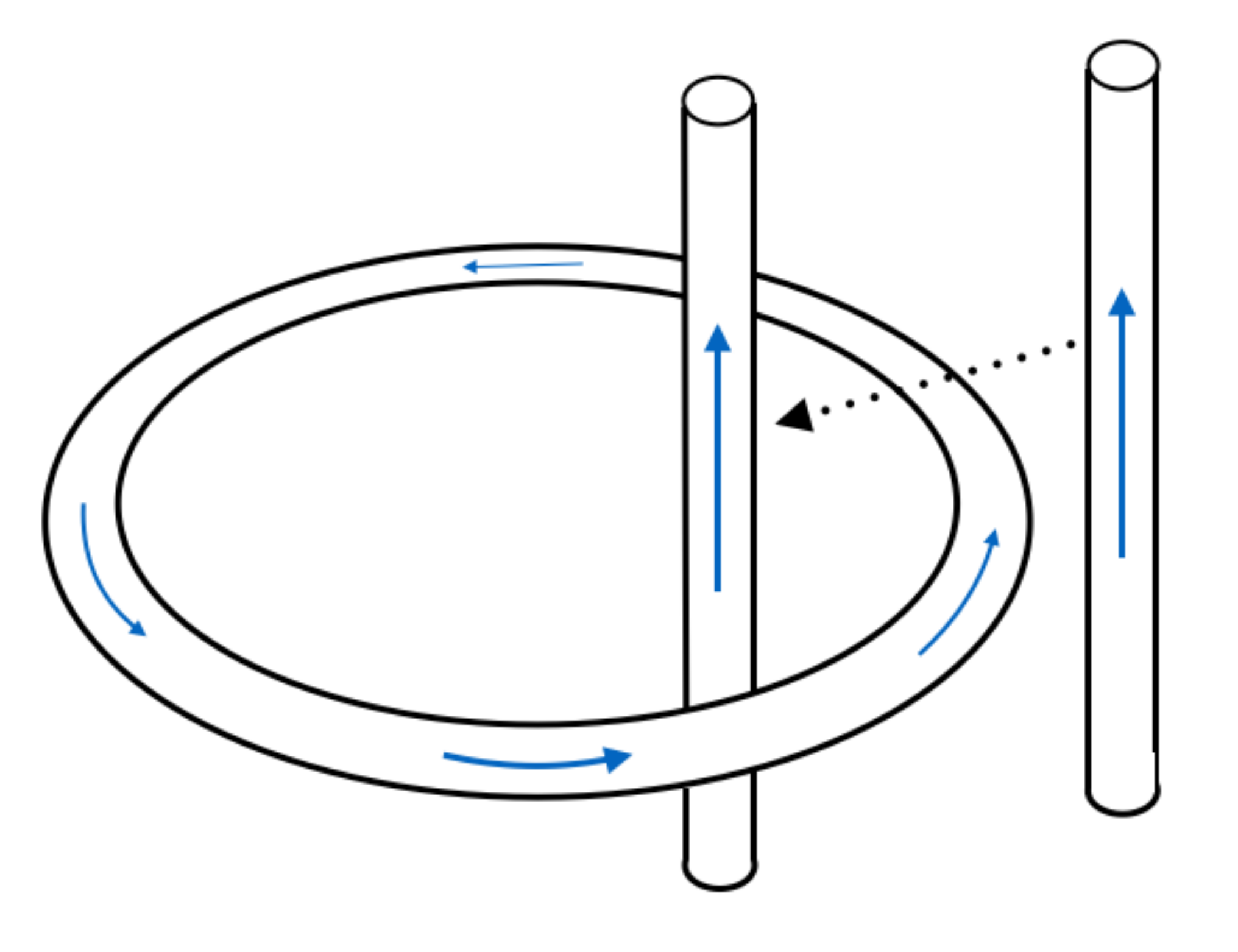}
\caption{
\label{fig:flux-in}
 (Color online)
A magnetic flux coming into a ring of another flux tube. 
 }
\end{figure}

The magnetic helicity can be changed either externally (flux
reconnection) or internally (flux twist). 
Consider first the topology change of the flux tubes by a magnetic
reconnection, as shown in Fig.~\ref{fig:flux-in}. The reconnection
leads to the change of the magnetic flux $\Phi$ flowing through the area
encircled by each of the tubes.
This change of the magnetic flux through Faraday's
induction generates an electric field parallel to the lines of the
magnetic field, 
\begin{equation}
\frac{d}{dt} \Phi = - \oint_{C_1} \bm E \cdot d \bm x , 
\end{equation}
where $\Phi$ is the magnetic flux that penetrates the loop $C_1$. The
change in the magnetic flux equals the flux contained inside the incoming tube,
$\Delta \Phi = \varphi_2 $.
Faraday's law allows us to write this change as 
\begin{equation}
 \Delta \Phi = \varphi_2 = - \oint_{C_1}
 \left[ \int_0^{\Delta t} \bm E(t) dt  \right] \cdot
 d \bm x,
 \label{eq:faraday}
\end{equation}
where $\Delta t$ is the amount of time needed for the reconnection.

In this derivation we will assume that the magnetic field $B$ is strong, such
that the magnetic length $(e B)^{-1/2}$ is small compared to the
thickness of the flux tubes. 
The chiral fermions are then localized on lowest Landau levels (LLLs),
and the system can be effectively treated as one dimensional.
The relevant degrees of freedom are the Landau zero modes. 
Since the LLLs are not degenerate in spin, the handedness of a fermion
is correlated with the direction of its motion (along or against the
direction of the magnetic field). Our discussion will be analogous to the
well-known description of the chiral anomaly in parallel electric and
magnetic fields developed in Refs.~\cite{Nielsen:1983rb,Witten:1984eb}.

The induced electric field changes the Fermi momenta of the left- and
right-handed fermions:
\begin{equation}
\dot{\bm p_F} = \pm e \bm E, 
\end{equation}
where the sign is plus (minus) for right- (left-)handed particles.
This change of the Fermi momentum implies 
that the particles (antiparticles) of right- (left-)handed species are produced. 
The change in the Fermi momentum due to the magnetic reconnection is thus 
\begin{equation}
 \Delta \bm p_F = \pm \int_0^{\Delta t} e \bm E(t) d t . 
\end{equation}
Integrating this over the circumference of the tube, 
\begin{equation}
\oint \Delta \bm p_F(s) \cdot d \bm x
  \equiv
 \int \Delta \bm p_F(s) \cdot \frac{d \bm x}{ds} ds
  =\pm  e \oint_{C_1}
  \left[ \int_0^{\Delta t} \bm E(t) dt  \right] 
  \cdot d \bm x  = \mp e \varphi_2 , 
\end{equation}
where $s$ is a variable that parametrizes the position along the flux,
and we have used Eq.~(\ref{eq:faraday}) in the last equality. 
If $e \varphi_2$ is positive,
there will be production of
right-handed antiparticles, because the Fermi energy decreases for
right-handed species. 
Since the density of states in (1+1) dimensions is given by $p_F / 2 \pi$, 
the number of produced antiparticles can be obtained as 
\begin{equation}
 \Delta \bar N_R =
   \frac{- \oint_{C_1} \Delta \bm p_F \cdot d \bm x}{2 \pi} \times \frac{e
   \varphi_1}{2 \pi }
   =  \frac{e^2 \varphi_1 \varphi_2 }{4 \pi^2}
   , 
\end{equation}
where $\varphi_1$ is the magnetic flux that forms the loop $C_1$, and 
$e \varphi_1 / 2 \pi$ is the Landau degeneracy factor describing the transverse density of states in the cross section of the tube.
On the other hand, for left-handed fermions, the Fermi energy increases,
which means that particles are created; their number is given
by 
\begin{equation}
 \Delta N_L
  =    \frac{e^2 \varphi_1 \varphi_2 }{4 \pi^2}. 
\end{equation}
The particle production thus leads to the generation of currents
of the left- and right-handed fermions, which
are given by the charge density times
velocity ($\pm 1$, respectively, for right- and left-handed currents) 
\footnote{The Fermi velocity (or the speed of light) is set to unity.}, 
\begin{equation}
 \oint_{C_1} \Delta \bm J_R \cdot d \bm x
  = (-e) \times \Delta \bar N_R
  = - \frac{e^3 \varphi_1 \varphi_2}{4 \pi^2 }, 
\end{equation}
where $(-e)$ is the charge of the antiparticle, and 
\begin{equation}
 \oint_{C_1} \Delta \bm J_L \cdot d \bm x
  =(-) \times e \times \Delta N_L 
  = - \frac{e^3 \varphi_1 \varphi_2}{4 \pi^2 }.
  \label{eq:jl}
\end{equation}
The minus sign in Eq.~(\ref{eq:jl}) comes from
the fact that the left-handed current flows in the opposite direction of 
the right-handed one. 
Therefore, the change in the total electric current is 
\begin{equation}
 \oint_{C_1} \Delta \bm J \cdot d \bm x
  =
 \oint_{C_1} \Delta \left[\bm J_R + \bm J_L \right]\cdot d \bm x
 = -\frac{e^3 \varphi_1 \varphi_2}{2 \pi^2 }.
 \label{eq:j-phi1-phi2}
\end{equation}
The flux coming into the loop $C_1$ is a part of another loop, $C_2$. 
One can easily convince oneself that the contribution to the integrated current
over $C_2$ is identical to that of $C_1$; therefore, we have 
\begin{equation}\label{prel_res}
\sum_{i\in \{1,2\} } \oint_{C_i} \Delta \bm J \cdot d \bm x
  =
\sum_{i\in \{1,2\} } \oint_{C_i}
  \Delta \left[\bm J_R + \bm J_L \right]\cdot d \bm x
  = -\frac{e^3}{ 2 \pi^2 }  \times 2 \varphi_1 \varphi_2. 
\end{equation}
Here we have factored out the quantity $2 \varphi_1 \varphi_2$, which is nothing but
the helicity change $\Delta \mathcal H$ in the process of switching
\cite{moffatt1990energy,Ricca1992,moffatt1992helicity}, as is shown
below. 
For two closed magnetic flux tubes, the magnetic helicity can be written as 
\begin{equation}
\mathcal H = \int  d^3 x\  \bm A \cdot \bm B = 2 n \varphi_1 \varphi_2, 
  \label{eq:mag-helicity}
\end{equation}
which can be illustrated as follows. 
When the tubes are very thin,
$\bm B$ is localized along two closed curves, and 
the magnetic field can be  written as 
\begin{equation}
\bm B(\bm x) =\varphi_1  \int \frac{d \bm X_1(s)}{ds} \delta(\bm x - \bm X_1(s))
 d s
  + \varphi_2  \int \frac{ d \bm X_2(s)}{ds} \delta(\bm x - \bm X_2(s))
  , 
\end{equation}
where $\bm X_{1,2}(s)$ are the coordinates of the two closed curves with a
parameter $s$. 
By plugging this expression into the definition of the magnetic
helicity~(\ref{eq:mag-helicity}), we get 
\begin{equation}
\mathcal H = \varphi_1 \oint_{C_1} \bm A \cdot d \bm X_1
  +
  \varphi_2 \oint_{C_2} \bm A \cdot d \bm X_2 . 
\end{equation}
The line integrals count the fluxes, given by the 
Gauss linking number $n$ between $C_1$ and $C_2$: 
\begin{equation}
 \oint_{C_1} \bm A \cdot d \bm X_1 = n \varphi_2 ,
  \quad
  \oint_{C_2} \bm A \cdot d \bm X_2 = n \varphi_1 .
\end{equation}
Thus, the magnetic helicity is expressed as 
\begin{equation}
\mathcal H = 2 n \, \varphi_1 \varphi_2  .
\end{equation}
Hence, the change in helicity, associated with the topology change of the
curves, is given by the change in the linking number, 
\begin{equation}
\Delta \mathcal H = 2 \left( \Delta n \right)  \, \varphi_1 \varphi_2
 .
 \label{eq:reconnection}
\end{equation}

Another way of changing the helicity is twisting. 
We can introduce a twist to a closed flux tube operationally,
as in Ref.~\cite{moffatt1990energy}. 
When a twist of angle $2 \pi \Delta n_0$ is introduced, 
the magnetic flux circled by a flux element $d \phi$ changes by
\begin{equation}
 \Delta \Phi =  \phi\ \Delta n_0  , 
\end{equation}
where $\phi$ is the flux inside.
%
This change of flux induces a CME current on the flux element $d \phi$,
the amount of which is given by (just as in the case of flux reconnection)
\begin{equation}
 \int_{d \phi} \Delta \bm J \cdot d \bm x =
  -2 \times  \frac{e^3}{2\pi^2} \phi\ \Delta n_0 d\phi
  =
  - \frac{e^3}{\pi^2} \phi\ \Delta n_0 d\phi
\end{equation}
where the factor 2 comes from the fact that the twisting of two flux
tubes leads to the generation of currents in both of them in an equal amount. 
The induced current in the whole flux tube is obtained by integrating
over the flux element, 
\begin{equation}
 \int_C \Delta \bm J \cdot d \bm x
  = 
\int d\phi 
 \int_{d \phi} \Delta \bm J \cdot d \bm x =
 - \frac{e^3}{2\pi^2} \varphi^2\ \Delta  n_0
 =
 - \frac{e^3}{2\pi^2}\ \Delta \mathcal H . 
 \label{eq:twist}
\end{equation}
Here we have used the fact that the increment in helicity
from twisting is $\Delta \mathcal H = \varphi^2\ 
\Delta  n_0$, where $\varphi$ is the magnetic flux of the tube being
twisted. 

Equations (\ref{eq:reconnection}) and (\ref{eq:twist}), 
combined with the expression for the current (\ref{prel_res}) derived above, yield 
our main result (\ref{main_res}).

\vskip0.3cm

A few comments are in order regarding the applicability of
Eq.~(\ref{main_res}). 

First, while deriving the formula, we have assumed that all of the flux
tubes are contained within the volume
of interest. The discussion can be naturally extended to the cases where  
the magnetic field is leaking from the volume.
Once the boundary condition is fixed 
between volume A and volume B, the helicity difference can be determined
with the knowledge of magnetic flux within volume A only (see Ref.~\cite{berger1984topological}),
and the derived formula applies to such cases as well.

Second, in the process of flux insertions,
Ohmic currents can also be generated through Faraday's law. 
Equation (\ref{main_res}) holds only for the CME contribution to the
current. The CME and Ohmic currents are different in nature and it is possible to
distinguish them. The Ohmic current dissipates, and
the CME current does not. If one waits long enough after a reconnection,
the Ohmic contribution dies off and only the CME current remains. 

Third, 
although our derivation of the formula is based on an assumption of the
LLL approximation and the homogeneous magnetic field, the derived equation
itself can hold on more general grounds, just as in the case of the
chiral magnetic effect. 
The CME can be explained in terms of the spectral flow of
the LLLs, but it can also be derived in hydrodynamics requiring the 
second law of thermodynamics \cite{Son:2009tf}. Likewise, we believe that there exist  
other ways of derivation using, for example, chiral kinetic theory. 
Still, let us discuss the applicability of assumptions we made 
in deriving the formula (\ref{main_res}). 
The discussed mechanism of the CME current generation resulting from the
change of the topology
of magnetic flux would operate in a plasma containing massless fermions, e.g. in the early Universe or in 
Dirac and Weyl semimetals. 
Thus, it seems reasonable to estimate the reconnection time scale
within magnetohydrodynamics. 
In magnetohydrodynamics, the time evolution of the magnetic field is
governed by the equation
\begin{equation}
 \p_t \bm B
  = \nabla \times (\bm v \times \bm B) + \frac{1}\sigma
  \nabla^2 \bm B , 
\end{equation}
where $\sigma$ is the Ohmic conductivity and $\bm v$ is the fluid
velocity. 
In order for a reconnection to occur, the
conductivity has to be finite, because, in the infinite conductivity
limit, the magnetic helicity is conserved and no
reconnections of magnetic field lines are present.  
The time scale that controls magnetic reconnections is given by
$
t_{\rm rec} \sim \sigma L^2
$, 
where $L$ is the typical length scale of the spatial inhomogeneity of
the magnetic field.
In order for the LLL approximation to be valid,
$t_{\rm rec}$ should be much longer than the inverse of the energy
difference between the Landau levels, $ t_{\rm Landau} \sim 1/\sqrt{ eB}$,
namely $t_{\rm rec } \gg t_{\rm Landau }$. This can be written as 
\begin{equation}
 \sigma L^2 \sqrt{eB} \gg 1 .
  \label{eq:cond-1}
\end{equation}
As for the assumption about the homogeneity of the magnetic field,
this is justified if the magnetic length $1/\sqrt{eB}$ is smaller than
$L$, from which we obtain another condition, 
\begin{equation}
 L \sqrt{eB} \gg 1. 
  \label{eq:cond-2}
\end{equation}
Hence, we expect that the scenario we describe in this Letter 
would be realized in a plasma with a massless fermion where the
conditions (\ref{eq:cond-1}) and (\ref{eq:cond-2}) are satisfied.

\vskip0.3cm
To summarize, we have demonstrated that the chiral magnetic current can
be generated without any initial chirality imbalance, by reconnections of
the magnetic flux. This current is entirely determined by the integer change
of the magnetic helicity and is, therefore, quantized.  Our result has a
number of implications -- for example, it will affect the evolution of
the magnetic helicity in chiral magnetohydrodynamics. Possible applications
include the quark-gluon plasma in heavy-ion collisions,
Dirac and Weyl semimetals, and primordial electroweak plasma produced after the big bang. 
\vskip0.5cm

\acknowledgements

This material is partially based upon work supported by the
U.S. Department of Energy, Office of Science, Office of Nuclear Physics,
under Contracts No. DEFG-88ER40388 (D.K.), DE-SC0012704 (Y.H., D.K., and
Y.Y.), and within the framework of the Beam Energy Scan Theory (BEST)
Topical Collaboration.
The work of Y.H. was partially supported by JSPS Research Fellowships
for Young Scientists.
D.K. also acknowledges the support of the Alexander von
Humboldt foundation and Le Studium foundation, Loire Valley, France,
for the support during the  ``Condensed matter physics meets relativistic
quantum field theory'' program.


\end{document}